\documentclass{ragtime} 
\usepackage{longtable}
\usepackage{subfigure}
\title[Trapped horizontal disk oscillations in Einstein's and Newton's gravity]
{Eigenmodes of trapped horizontal oscillations in accretion
 disks}
\author[ S. Khanna, 
         Z. Strzelecka,   
        B. Mishra and
        W. Klu\'zniak]
       { S. Khanna\at{1} 
         Z. Strzelecka\at{2} 
        B. Mishra\at{3}
        W. Klu\'zniak\at[]{3} \\
        \ins{1}Leiden Observatory, Leiden University, PO Box 9513,\\
~~2300 RA Leiden, The Netherlands \\
        \ins{2}Imperial College London, Department of Physics,
 Blackett Laboratory,\\ ~~Prince Consort Road, London SW7 2BB, United Kingdom \\
        \ins{3}Nicolaus Copernicus Astronomical Center, Bartycka 18, 00-716 Warsaw, Poland }
\coentry{Z. Stuchl\'{\i}k, J. Kov\'a\v{r} and J. Cimrman}%
       {Equilibrium of spinning test particles in equatorial plane\par
        of Kerr--de~Sitter spacetimes}

\providecommand{\dif}{\mathrm{d}} 

\begin{document}

\begin{abstract}
We present eigenfrequencies and eigenfunctions of trapped
acoustic-inertial oscillations of thin accretion disks
for a Schwarzschild black hole and a rapidly rotating Newtonian star
(a Maclaurin spheroid).
The results are derived in the formalism of  \citet{nowak1991}
with the assumption that the oscillatory motion is parallel to the midplane
of the disk. 
The first four radial modes for each of five azimuthal modes
($m=0$ through $m=4$) are presented. The frequencies and wavefunctions
of the lowest modes may be  accurately approximated by Airy's function.
\end{abstract}

\begin{keywords}
Relativistic stars: black holes, structure stability and oscillations,
relativity and gravitation, accretion disks, hydrodynamics
\end{keywords}
\section{Trapped modes}

\citet{fukato1980} showed that acoustic-inertial modes 
may be trapped in the inner parts of an accretion disk.
This occurs when the (radial) epicyclic frequency $\kappa$ has a maximum,
as is the case in the Schwarzschild metric of general relativity (GR)
considered by the authors.
\citet{okazaki1987,kato89} and \citet{nowak1991,nowak1992}
consider a model of a black hole
accretion disk in hydrostatic equilibrium, and
derive a dispersion relation for modes with $n=0,1,2,3..$
nodes along the $z$-axis (the symmetry axis of the disk).
The trapping occurs for oscillation frequencies below the maximum
of the epicyclic
frequency $\omega<\kappa_{\rm max}$.
Here $\omega(r)=m\Omega(r)+\sigma$ is
the frequency in the frame co-rotating with the fluid 
(at angular frequency $\Omega$), $m$ is the azimuthal mode number,
$\sigma$ is the eigenfrequency of the mode, and 
$\kappa^2=(2\Omega/r)\dif (r^2\Omega)/\dif r$.
The $n=0$ modes will be trapped between the inner edge of the disk,
close to the ISCO at $\kappa(r_{\rm ms})=0$,
and the lowest radius $r$ satisfying $\omega(r)=\kappa(r)$,
while for $n=1$ trapping occurs close to the maximum of $\kappa$,
between those two radii at which  $\omega=\kappa$.
Further discussion can be found in the textbook by \citet{bluebook}.
In this contribution we only consider the $n=0$ trapped modes.

Currently, the main interest in disk oscillations is related to the
observed frequencies in the X-ray flux from black hole and neutron
star systems \citep[for a review see][]{vdk00}. For black hole disks
the modes thought to be offering the most promising explanation
\citep{bob2001} of the highest observed frequencies  are the
$g$-modes and $c$-modes, investigated in full GR by \citet{bob1997,silber01},
although a different explanation seems to be required for the
observed 3:2 ratio of the highest frequencies in the microquasars
\citep{2001A&A...374L..19A,2004AIPC..714..379K,2005A&A...436....1T}. 
Thus, the modes investigated here are not prime candidates for
a theoretical counterpart to the observed high frequency QPOs
(quasi-periodic oscillations) in black hole systems. However,
similar phenomena are observed in white dwarf systems
\citep{woudt02}, and while their harmonic content may be explained
by a resonance \citep{2005A&A...440L..25K}, the origin of
the observed frequencies remains obscure. For this reason
we would like to discuss disk oscillations in a framework
valid equally in a GR and non-GR context. 

\begin{figure}[b]
\begin{center}
\includegraphics[width=\linewidth,height=.7\linewidth]{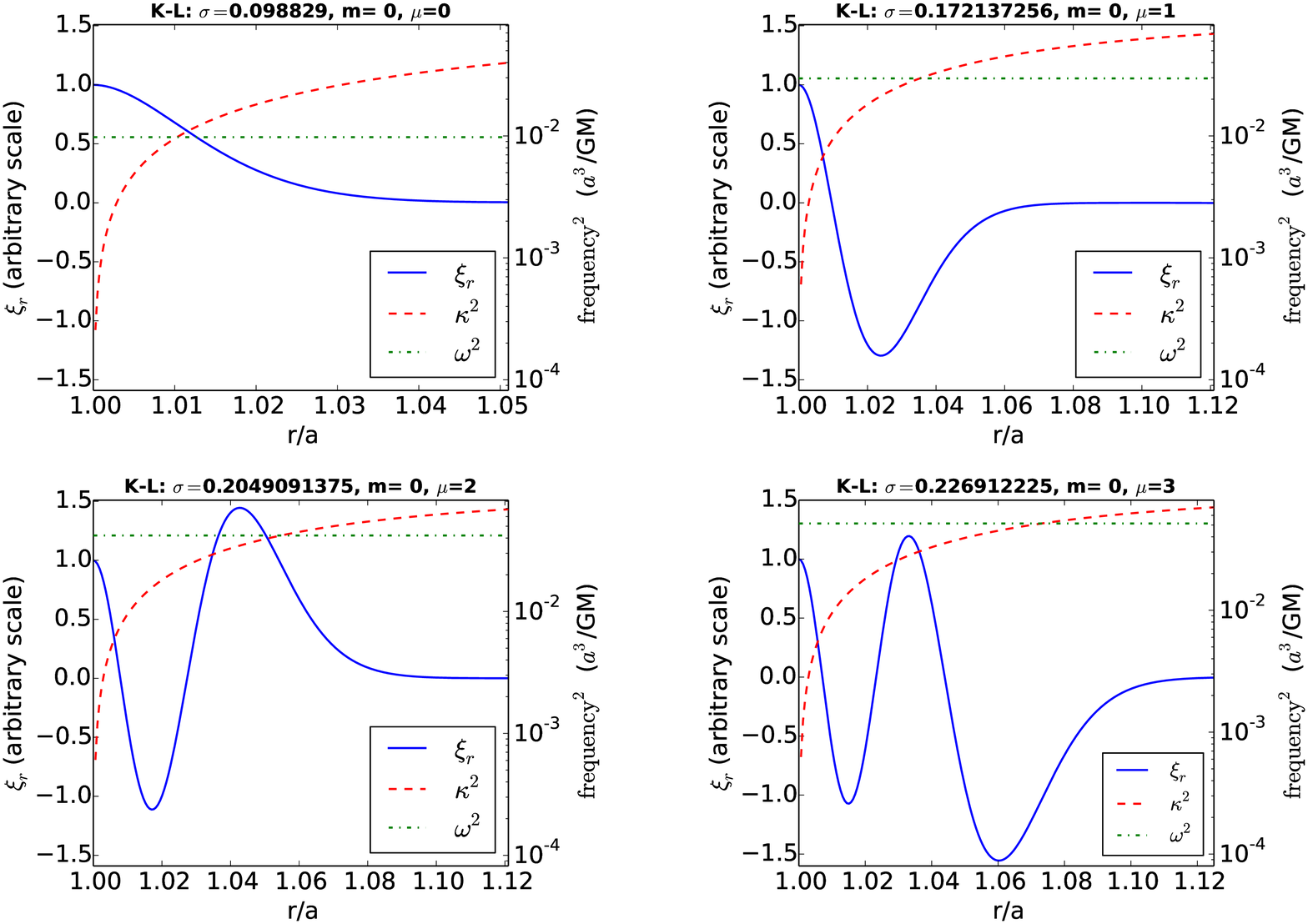}
\end{center}
\par\vspace{-1ex}\par
\caption{The fundamental and the first three radial overtones for $m=0$
 trapped horizontal oscillations of a thin
($H/a=10^{-3}$) accretion disk for the potential of Eq.~(\ref{kl}).
 Plotted are the  wavefunction:
solid  (blue) line 
(arbitrary normalization, left scale);
$\omega^2(r)/\Omega^2(r_{\rm ms})$: dashed-dotted (green) line
and $\kappa^2(r)/\Omega^2(r_{\rm ms})$:
dashed (red) line (logarithmic scale, right).}
\label{Fig:1}
\end{figure}

\begin{figure}[b]
\begin{center}
\includegraphics[width=\linewidth,height=.7\linewidth]{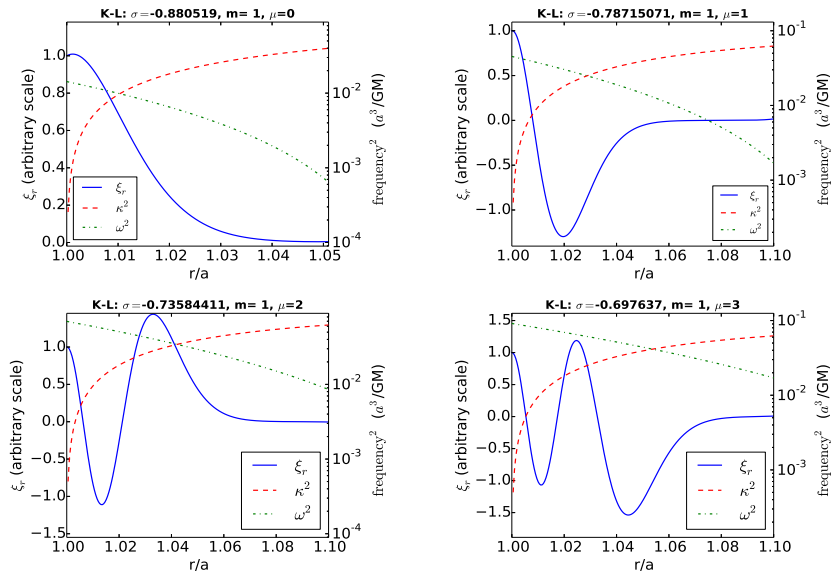}
\end{center}
\par\vspace{-1ex}\par
\caption{Same as Fig.~\ref{Fig:1}, but for $m=1$.}
\label{klm1}
\end{figure}

\section{Equation of motion and the boundary condition}

We will be closely following the approach of \citet{nowak1991}
who describe perturbations
with a Lagrangian displacement vector in cylindrical coordinates
$(\xi_{*}^r,\xi_{*}^{\phi},\xi_{*}^z)
=(\xi^r,\xi^{\phi},\xi^z)\exp[i(m\phi+\sigma t)]$
 in the formalism of  \citet{friedman78}, and
 show that in the WKB approximation
the azimuthal component of the equation of perturbed
motion for thin disks reduces to
$\xi ^{\phi} = 2i \big({\Omega}/{\omega}) \xi ^{r}$.
In this contribution we assume horizontal motion,
implying that $\xi_{*}^z\equiv 0$
 and ${\partial{\xi_{*}^r}}/{\partial z}\equiv0$.
In terms of $\Psi(r)\equiv \sqrt{\gamma P r}\,\xi^r(r)$
the remaining component of the equation of motion then gives
\begin{equation}
\frac{\dif ^{2} \Psi}{\dif r^{2}} +
 \frac{(\omega ^{2} - \kappa ^{2})}{c_{\rm s} ^{2}}\Psi = 0,
\label{bob's}
\end{equation}
where $c_{\rm s}^2=\gamma P/\rho$ is the speed of sound squared;
the boundary condition is that the Lagrangian perturbation of pressure
vanishes at the unperturbed boundary, 
$\Delta P\equiv \gamma P \nabla \xi_*=0$, which reduces to 
$$
\frac{1}{r} \frac{\partial}{\partial r} (r \xi _{*} ^{r}) 
+ \frac{1}{r} \frac{\partial}{\partial \phi} ( \xi _{*} ^{\phi})  = 0
$$
assuming that $P\ne0$
\citep{nowak1991}. Neglecting derivatives of $P$ this gives our
final boundary condition at the inner edge, at $r=a$,
which we will take to be at the marginally stable orbit (ISCO)
at $a=r_{\rm ms}$,
$$
\frac{\dif \Psi}{\dif r} =
 - \frac{\Psi}{2r} \big(1 - 4m {\Omega}/{\omega} \big) .
$$
In dimensionless form, with $r=a(1+x)$,
$\tilde\omega(x) =\omega(r)/\Omega(a)$, 
$\tilde\kappa(x) =\kappa(r)/\Omega(a)$,
$\tilde\sigma =\sigma/\Omega(a)$,
and $c_{\rm s}=H\Omega(a)$,
 the perturbation
(wave) equation takes the form
\begin{equation}
\frac{\dif ^{2} \Psi}{\dif x^{2}} +
  \left(\frac{a}{H} \right)^2\left(\tilde\omega ^{2}
 - \tilde\kappa ^{2}\right)\Psi = 0,
\label{prime}
\end{equation}
with the boundary condition at $x=0$
\begin{equation}
\frac{\dif \Psi}{\dif x} =
 - \frac{\Psi}{2} \big(1 - 4m/{\tilde\omega} \big) .
\label{bc}
\end{equation}
In the last equation $\tilde\omega=\tilde\sigma +m$.
Recall that in general $\tilde\omega(x)=\tilde\sigma + m\Omega(r)/\Omega(a)$.
\begin{figure}[b]
\begin{center}
\includegraphics[width=\linewidth,height=.7\linewidth]{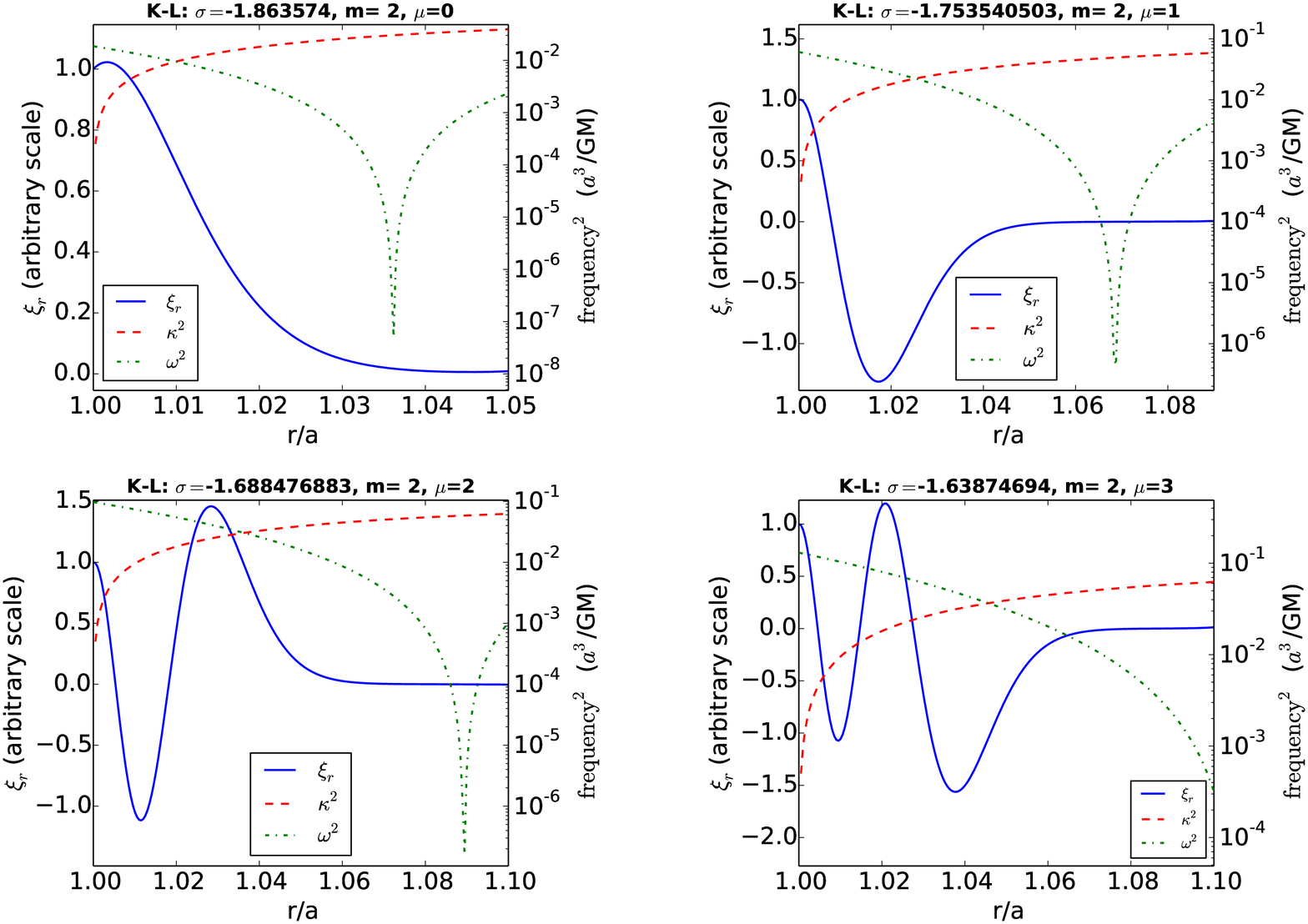}
\end{center}
\par\vspace{-1ex}\par
\caption{Same as Fig.~\ref{Fig:1}, but for $m=2$.}
\label{klm2}
\end{figure}

In this contribution we are providing an atlas of
eigenfrequencies and eigenfunctions for the fundamentals and the first
three radial overtones of horizontal disk oscillations
(labeled with the number of radial nodes,
 $\mu=0,1,2,3$) for $m=0,1,2,3,4$.
\section{
Models of a Schwarzschild black hole}
\begin{figure}[b]
\begin{center}
\includegraphics[width=\linewidth,height=.7\linewidth]{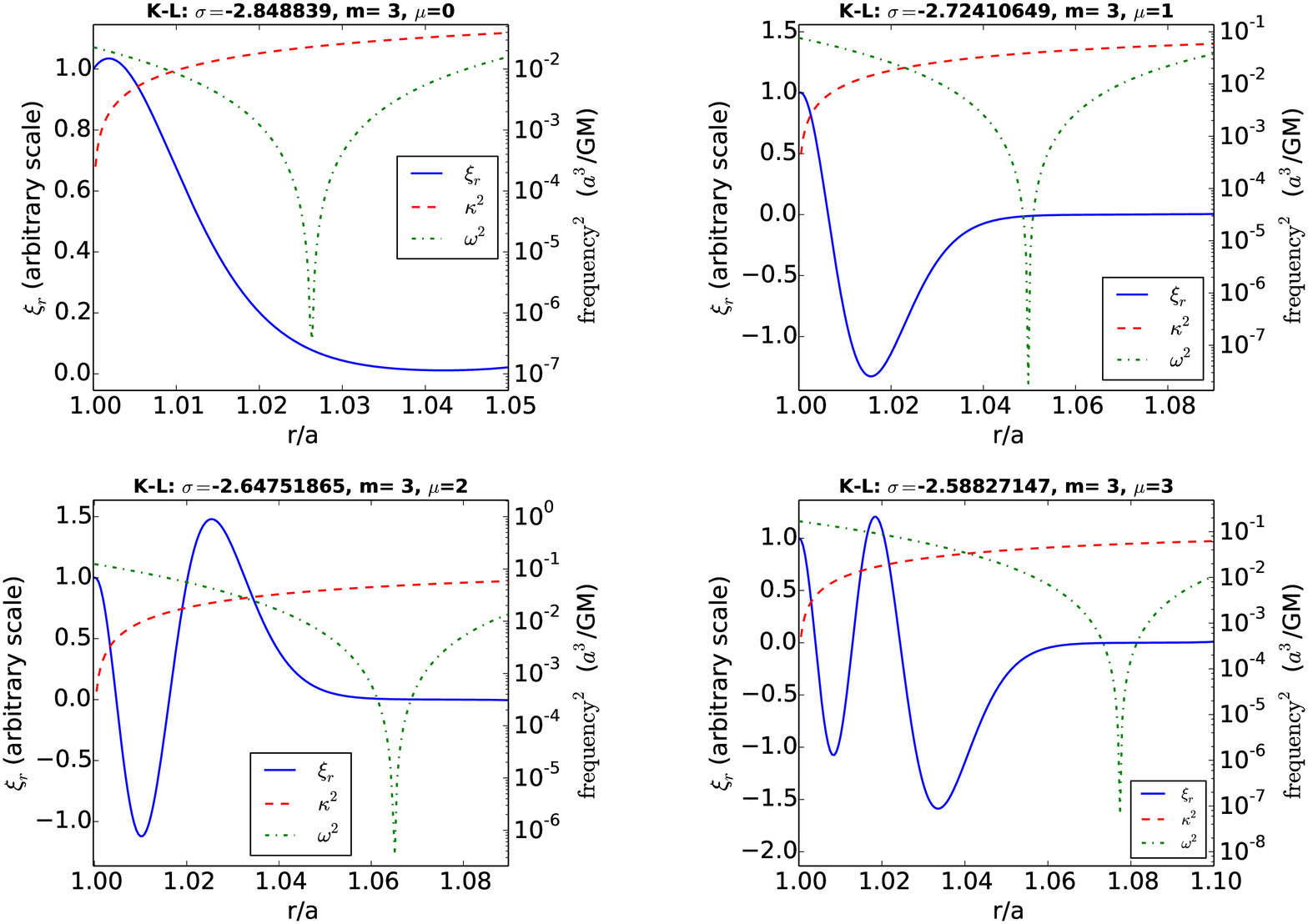}
\end{center}
\par\vspace{-1ex}\par
\caption{Same as Fig.~\ref{Fig:1}, but for $m=3$.}
\label{klm3}
\end{figure}

Bohdan Paczy\'nski showed that it is possible to capture
essential qualitative features of motion in the Schwarzschild metric
in a Newtonian model with a simple pseudo-potential $\Phi(r)=-GM/(r-2r_g)$
\citep{PW1980}, with $r_g=GM/c^2$. 
\citet{nowak1991} found the eigenfrequencies and eigenfunctions
of Eq.~(\ref{bob's}) for the fundamental oscillations with $m=0$, and $m=2$,
using values of $\kappa^2(r)$ following from their own pseudo-potential
$\Phi(r)=-(GM/r)[1-6r_g/r+12(r_g/r)^2]$.

\begin{figure}[b]
\begin{center}
\includegraphics[width=\linewidth,height=.7\linewidth]{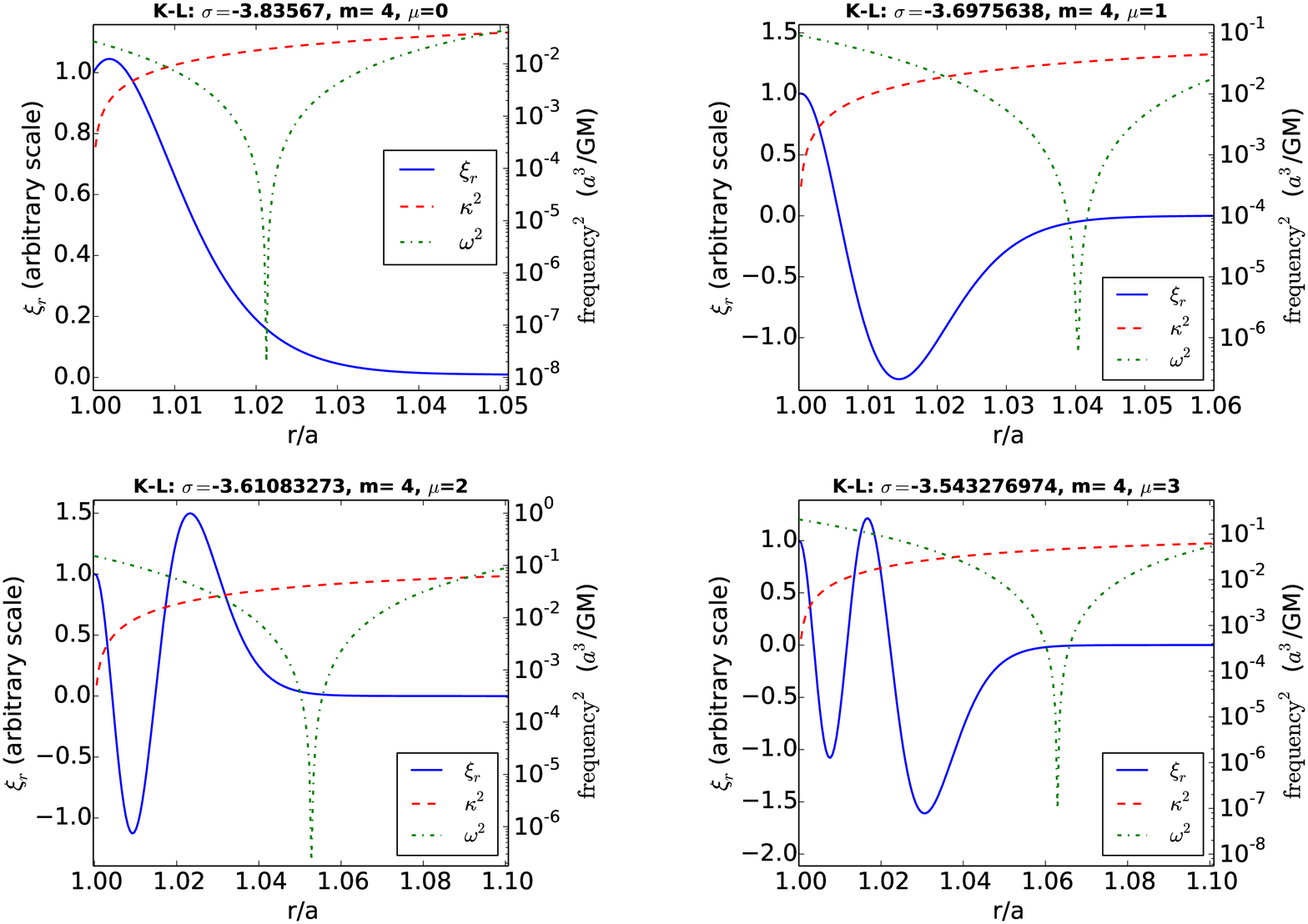}
\end{center}
\par\vspace{-1ex}\par
\caption{Same as Fig.~\ref{Fig:1}, but for $m=4$.}
\label{klm4}
\end{figure}

 Here, we model the Schwarzschild metric
with a Newtonian pseudo-potential designed expressly to reproduce the
Schwarzschild ratio of $\kappa^2(r)/\Omega^2(r)=1-6r_g/r$:
\begin{equation}
\Phi_{\rm KL}(r)=-(c^2/6)\exp(6r_g/r-1).
\label{kl}
\end{equation}
As we are only interested in the inner parts of an accretion disk,
we have dropped an additive constant. We have also renormalized the
original form of the potential \citep{kl2002} by a factor of $1/e$
to guarantee the correct value of $\Omega(r_{\rm ms})$.
The angular frequency of orbital motion follows from
   $\Omega^2(r)=r^{-1}\partial\Phi_{\rm KL} /\partial r$
and, as for the other two potentials,
the marginally stable orbit comes out to be at $r_{\rm ms}=6GM/c^2$.
We have numerically solved the eigenvalue problem given by
 Eqs.~(\ref{prime}), (\ref{bc}), for $H/a=10^{-3}$.
The equations being linear in $\Psi$,
we normalize the wavefunction to unity at the inner edge of the disk:
 $\Psi(r_{\rm ms})=1$.
Fig.~\ref{Fig:1} presents the eigenfrequencies $\sigma$
 and the eigenfunctions $\Psi(r)$ for $m=0$ and $\mu=0,1,2,3$,
while Figs.~\ref{klm1}, \ref{klm2}, \ref{klm3}, \ref{klm4} present the same
quantities, as well as $\tilde\omega^2$, for $m=1,2,3,4$, respectively.

\section{Essentials of acoustic-inertial oscillations}
\subsection{Wave equation}
\begin{figure}[b]
\begin{center}
\includegraphics[width=\linewidth,height=.7\linewidth]{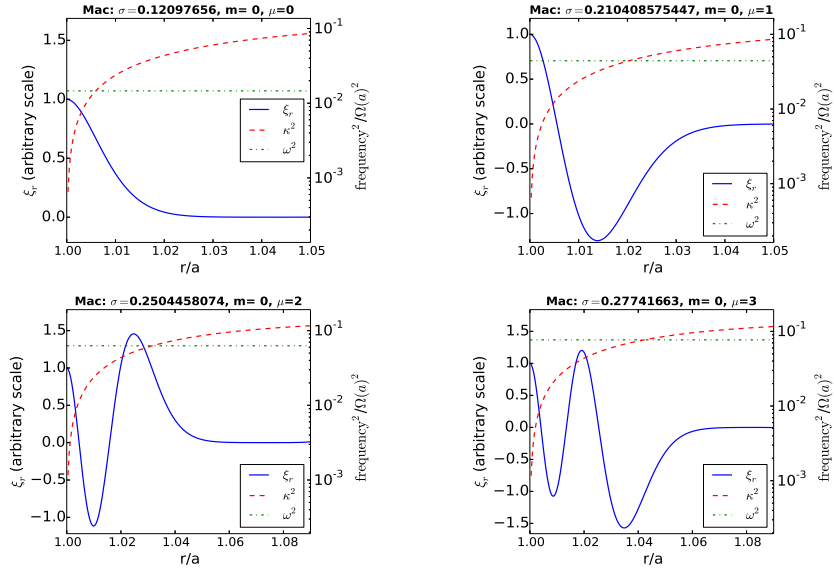}
\end{center}
\par\vspace{-1ex}\par
\caption{Same as Fig.~\ref{Fig:1}, but for the potential
of a Maclaurin spheroid with ellipticity $e = 0.834 583 178$.}
\label{mac0}
\end{figure}

\begin{figure}
\includegraphics[width=.48\linewidth,height=.33\linewidth]{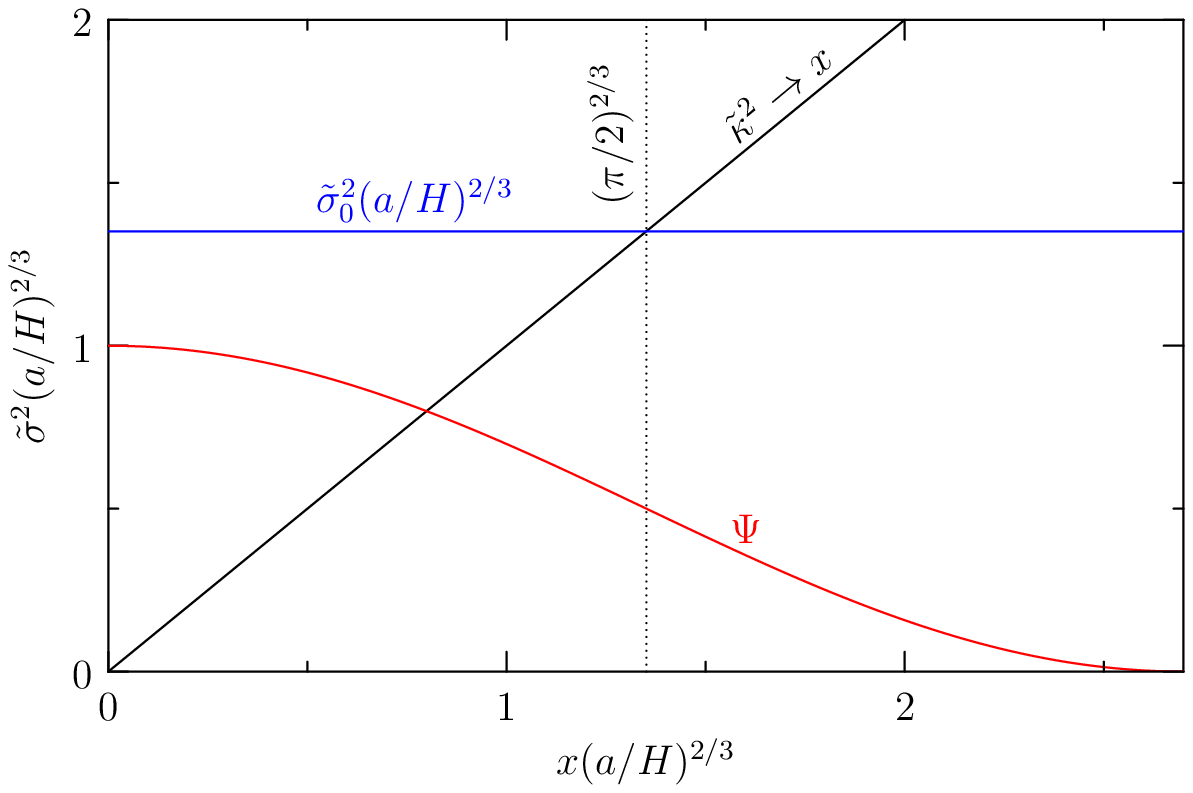}
\includegraphics[width=.52\linewidth,height=.35\linewidth]{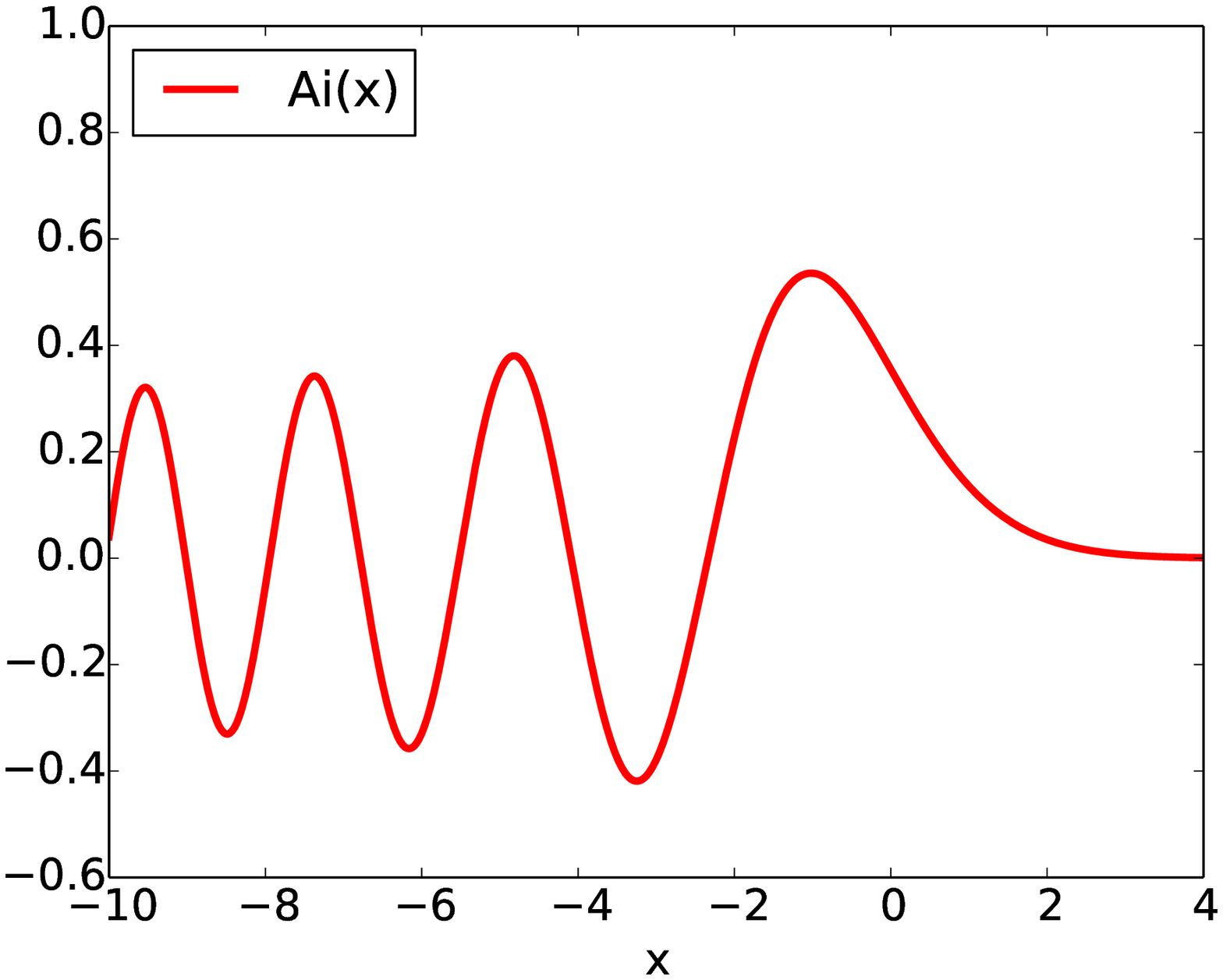}
\hspace{5mm}
\par\vspace{-1ex}\par
\caption{{\sl Left Panel:}
 A simple approximation to the eigenfrequency $\tilde\sigma_0$
of the axisymmetric fundamental mode ($m=0, \mu=0$):
 a quarter wavelength fits in the inner region of the disk
($\tilde\sigma^2_0\ge\tilde\kappa^2$) before the wave becomes evanescent.
The vertical dotted line indicates the value of
$x$ at which $\tilde\sigma^2_0$ and $x$ must intersect for this condition
to be met:   $\tilde\sigma^2_0=\tilde\kappa^2(x_0)\approx x_0$, with
$x_0=(\pi/2)^{2/3}\cdot(H/a)^{2/3}$.
The diagonal solid (black) line corresponds to the linear approximation
$\tilde \kappa^2(x)= x +{\cal O}(x^2)$, which is valid for the potential
of Eq.~(\ref{kl}). 
See Eq.~(\ref{model0}) and Section \ref{modele} for details
\newline{\sl Right Panel:} A better approximation is obtained
 from the location of the
extrema of Airy's function Ai($X$). Note that the shape of Ai($X$)
closely resembles the shape of the numerically found $\Psi(x)$.}
\label{model}
\end{figure}

Dust orbiting an axially symmetric gravitating body in its equatorial plane
($z=0$) would settle in stable circular orbits. The orbit of
each dust particle being stable,
it corresponds to ``rest''
(we are only concerned with radial motion in this section)
at a fixed radial distance from the center of the body
in the minimum of the effective potential, 
$V(r,z)=\Phi(r,z)+l^2/(2r^2)$,
 $l\equiv r^2\Omega(r)$ being the conserved angular momentum
of a given particle, and $\Phi$ the gravitational potential of the
body, both per unit mass.
 Consider small radial perturbations $\delta r$ of
motion of a dust disk (such as the rings of Saturn).
Neglecting particle collisions, the perturbed dust would be executing
radial harmonic (epicyclic) motion with respect of the stable orbits.
The square of the frequency of this radial motion, 
$\kappa^2=\partial\,^2V/\partial\,^2 r$ corresponds 
to the strength of the restoring
force per unit mass: $-\kappa^2\psi_*$ (if we denote the radial displacement
$\delta r=\psi_*$).
If the dust disk is replaced by a fluid, there will be an additional
restoring force corresponding to pressure perturbations.

It is well known that sound waves in a homogeneous medium can be described
by a harmonic function both in space and in time, with a constant and uniform
amplitude if attenuation is neglected. Thus, the acoustic displacement 
of the fluid satisfies both a wave equation
\begin{equation}
\frac{\partial ^{2} \psi_*}{\partial y^{2}} +
{k_{\rm s} ^{2}} \psi_* = 0,
\label{wave}
\end{equation}
and an oscillator equation
\begin{equation}
\frac{\partial ^{2} \psi_*}{\partial t^{2}} +
{\omega_{\rm s} ^{2}} \psi_* = 0,
\end{equation}
corresponding to a restoring force $-\omega_{\rm s}^2\psi_*$.
The frequency of the sound wave is related to the wave vector
through the linear dispersion relation
\begin{equation}
k_{\rm s}^2=\omega_{\rm s}^2/c_{\rm s}^2.
\label{vector}
\end{equation}

Clearly, taking into account in the oscillator equation
both the ``inertial'' (epicyclic) and the acoustic
restoring forces,
and neglecting for the moment the difference between the cylindrical
co-ordinate $r$ and the Cartesian  co-ordinate $y$, 
the acoustic-inertial displacement of the fluid
can be described by a displacement $\psi_*(y,t)=\psi(y)\exp(i\omega t)$,
with $\omega^2=\kappa^2+\omega_{\rm s}^2$
\citep[or, in the form written down by][, $\omega^2=
\kappa^2+k_{\rm s}^2c_{\rm s}^2$]{binney87}. 
Substituting this new dispersion relation
into Eq.~(\ref{vector}), we see that Eq.~(\ref{wave}) takes the form
\begin{equation}
\frac{\dif ^{2} \psi}{\dif y^{2}} +
\frac{\omega^2-\kappa^2}{c_{\rm s}^2} \psi= 0.
\label{bis}
\end{equation}
Remarkably, this is the same equation that was rigorously derived
by \citet{nowak1991}, i.e., Eq.~(\ref{bob's}). In the remainder of this paper
we will be discussing numerical solutions of its dimensionless version,
Eq.~(\ref{prime}), subject to the boundary condition Eq.~(\ref{bc}),
 for a thin disk ($H/a=0.001$) in two different models of the gravitating body,
 i.e., for two different epicyclic frequencies $\kappa(r)$.
\begin{figure}[b]
\begin{center}
\includegraphics[width=\linewidth,height=.7\linewidth]{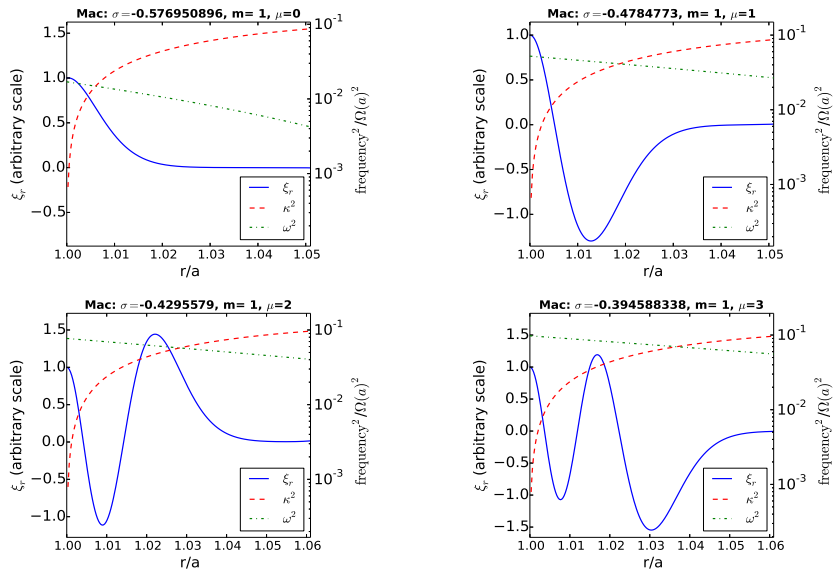}
\end{center}
\par\vspace{-1ex}\par
\caption{Same as Fig.~\ref{mac0}, but for $m=1$.}
\label{mac1}
\end{figure}

\subsection{Estimates of the eigenfrequencies}
\label{modele}
It is possible to understand the values of the eigenfrequencies $\sigma$
and the shape of the wavefunctions in a simple model of
 Eq.~(\ref{prime}). For axially symmetric modes, $m=0$ and hence
$\omega=\sigma$.
The wave equation has oscillatory solutions for $\omega^2>\kappa^2$,
while the wave is evanescent for $\omega^2<\kappa^2$. Thus the mode
is trapped between $a=r_{\rm ms}$ (i.e., $x=0$) and $r=r_0$ such that
$\sigma^2=\kappa^2(r_0)$ (Fig.~\ref{model}).

As $\sigma^2<<\kappa^2_{\rm max}$ for the fundamental mode
and $\kappa^2(r_{\rm ms})=0$
we can model $\kappa^2$ with a linear approximation \citep{nowak1991}, which
for the potential of Eq.~(\ref{kl}) has the simple 
form $\tilde\kappa^2=x$. Thus, the wave becomes evanescent at 
$r_0/a-1=x_0\approx \tilde\sigma^2$.
 We can take the boundary condition on the wave to correspond
to that of a banner flapping in the wind, with a crest at the edge
[of the disk ($x=0$)] and a node close to $x_0$.
 Perhaps a quarter wavelength of a sinusoid
between $x=0$ and $x=x_0$ is a fair approximation \citep{fukato1980}.

With the above approximations, we have $ax_0=\lambda/4$,
and $\tilde\sigma^2=x_0$. Now, $k=2\pi/\lambda\approx\sigma/c_{\rm s}$
so  $\lambda/4\approx \pi c_{\rm s}/(2\sigma)=\pi H\Omega/(2\sigma)$.
Recall that $\tilde\sigma=\sigma/\Omega(a)$. Finally, we obtain
$\tilde\sigma^3=\pi H/(2a)$,
yielding 
\begin{equation}
\tilde\sigma\approx\left(\frac{\pi H}{2a}\right)^{1/3}\approx1.16
\left(\frac{H}{a}\right)^{1/3}.
\label{model0}
\end{equation}
For $H/a=10^{-3}$ this yields $\tilde\sigma\approx0.116$, while the numerically
obtained value for the correct functional form of $\kappa^2$
is $\tilde\sigma_0\approx0.0988$. Thus, this crude estimate of
the eigenfrequency is off by less than $20\%$. However, as we will see
directly below,
we have obtained the correct scaling of the eigenfrequency
with the dimensionless thickness
of the disk \citep{fukato1980}.

A more accurate estimate of the eigenfrequency can be obtained
by noting that in the linear approximation to $\kappa^2$
(which for the potential of Eq.~[\ref{kl}] is simply
$ \tilde\kappa^2=x$),
 Eq.~(\ref{prime}) corresponds to Airy's equation
 \citep{nowak1991}. Indeed, with the substitution
$X=(x-\tilde\sigma^2)(a/H)^{2/3}$, Eq.~(\ref{prime}) becomes 
$
{\dif^2\Psi}/{\dif X^2}=X\Psi,
$
with the Airy function as the solution: $\Psi(X)={\rm Ai}(X) $.
In the exact waveforms of Fig.~\ref{Fig:1}, one can recognize the shape
of Airy's function, to a good accuracy.
The (implicit) eigenvalues $\tilde\sigma$ can now be found directly
from the boundary condition, Eq.~(\ref{bc}), in the form
$$
\frac{1}{\Psi}\frac{\dif\Psi}{\dif X}=
\left(\frac{H}{a}\right)^{2/3}\cdot\frac{1}{2} \big(4m/{\tilde\omega-1} \big).
$$
Now, for $H<<a$, the boundary condition (at $x=0$) becomes
${\dif\log\Psi}/{\dif X}<<1$,
i.e., it is approximately that $X$ corresponds to one of those $X_\mu$
for which Ai($X_\mu$) has an extremum, $\dif{\rm Ai}/\dif X|_{X_\mu}=0$.
 Thus, 
$\tilde\sigma^2_\mu\approx-(H/a)^{2/3}X_\mu$, $\mu=0,1,2,3...$.
We can compare
these approximate eigenfrequencies (Table~\ref{Airy}) with the numerically
found eigenvalues for the correct form of $\kappa^2$. For the fundamental
the agreement is quite good, but the accuracy of the Airy approximation
gradually degrades as $\sigma_\mu$ approaches the value $\kappa_{\rm max}$.

\begin{longtable}{c|c|c|c|c}
\caption{Exact and approximate eigenvalues of Eqs.~(\ref{prime}),~(\ref{bc})}\\
\hline
$m=0,\,\, H/a=0.001$& $\mu=0$ & $\mu=1$ &$\mu=2$  & $\mu=3$\\
\hline
&&&&\\
$\tilde\sigma_\mu$ for $\kappa^2$ of Eq.~(\ref{kl})
 & 0.0988... & 0.172... & 0.205... & 0.227... \\
Airy approx.: $\sqrt{-0.01X_\mu}$
 & 0.101... & 0.180... & 0.229... & 0.248... \\
Accuracy of approximation
 & 2\% & 5\% & 7\% & 9\% \\
\hline
\label{Airy}
\end{longtable}
We thank Mr. Luca Giussani for providing us with the values of Airy's
extrema.

%

\section{Trapped oscillations in an accretion disk around a Maclaurin spheroid}
\label{mac}
In previous sections, following \citet{nowak1991}
 we were discussing the trapped acoustic-inertial oscillations of 
a pseudo-Newtonian model of an accretion disk around a Schwarzschild
black hole. Interestingly, the same trapping
phenomenon occurs in strictly Newtonian
gravity, for disks orbiting sufficiently oblate bodies.
\citet{kbg-r01}, and \citet{zdunik01} pointed out that oblateness
of a gravitating body can destabilize orbits close to it,
while \citet{amster02} showed that the marginally stable orbit
exists in the Newtonian potential of classic Maclaurin spheroids
for a sufficiently large ellipticity of the spheroid, i.e., a sufficiently
large rotation rate of the spheroid.
\citet{krosinska2013} give explicit expressions for the angular velocity
in circular orbits and for the corresponding epicyclic frequencies
as a function of orbital radius and the ellipticity of the Maclaurin spheroid.
 \citet{mateusz2014} compare these analytic expressions
with exact numerical solutions (in GR) of rapidly rotating quark stars,
while \citet{mishra2014} give accretion disk solutions in the gravitational
field of Maclaurin spheroids, which are reminiscent of the \citet{ss73}
black hole accretion disks.

\begin{figure}[b]
\begin{center}
\includegraphics[width=\linewidth,height=.7\linewidth]{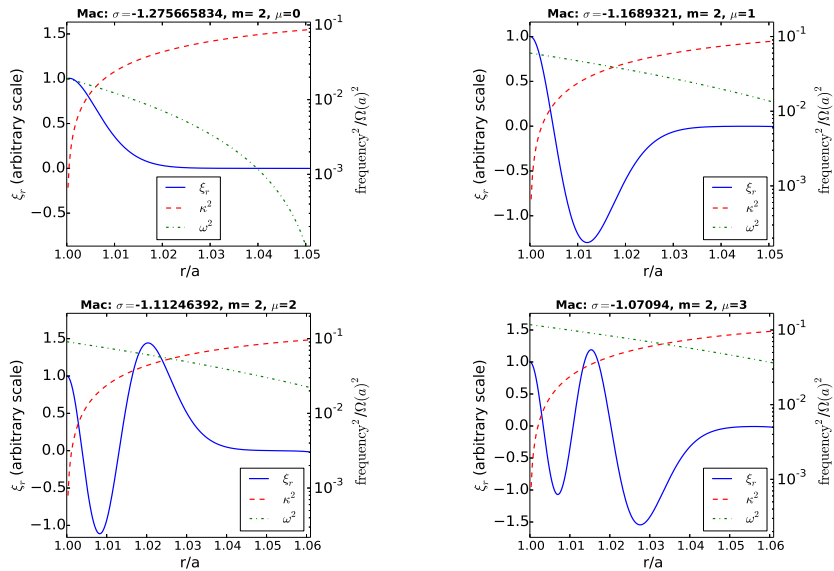}
\end{center}
\par\vspace{-1ex}\par
\caption{Same as Fig.~\ref{mac0}, but for $m=2$.}
\label{mac2}
\end{figure}

\begin{figure}[b]
\begin{center}
\includegraphics[width=\linewidth,height=.7\linewidth]{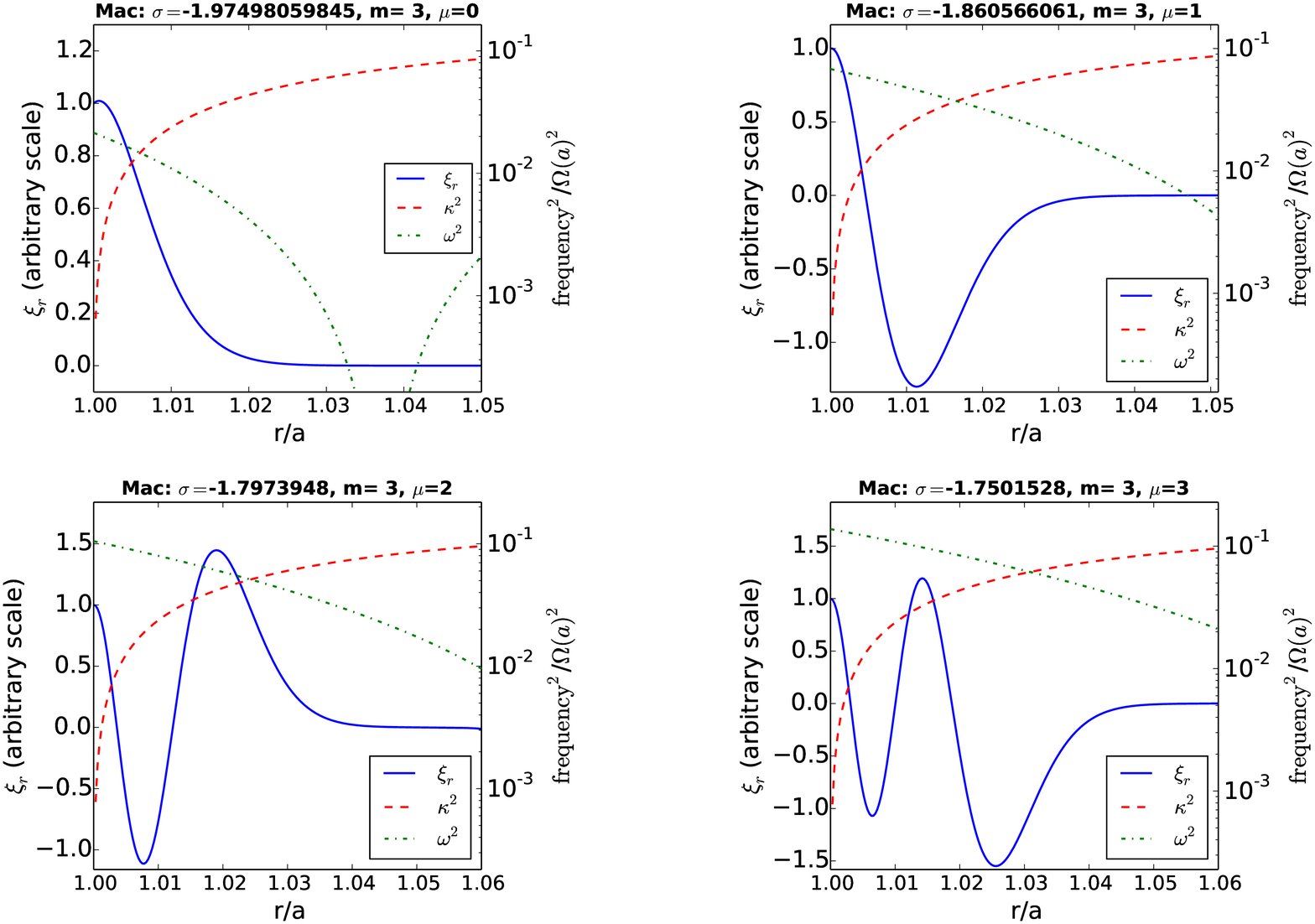}
\end{center}
\par\vspace{-1ex}\par
\caption{Same as Fig.~\ref{mac0}, but for $m=3$.}
\label{mac3}
\end{figure}

\begin{figure}[b]
\begin{center}
\includegraphics[width=\linewidth,height=.7\linewidth]{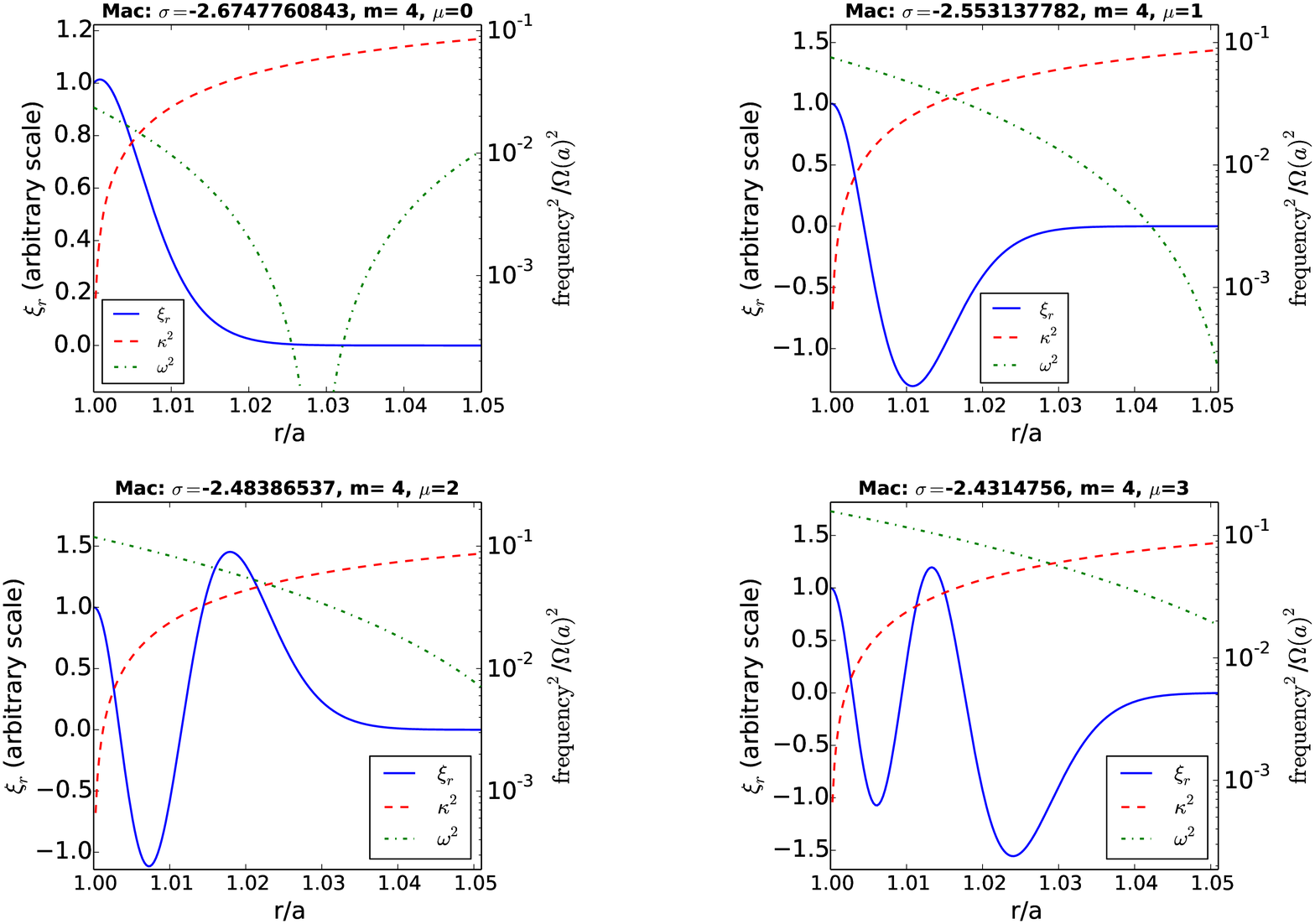}
\end{center}
\par\vspace{-1ex}\par
\caption{Same as Fig.~\ref{mac0}, but for $m=4$.}
\label{mac4}
\end{figure}

Without further ado, we are presenting the eigenfrequencies and eigenvalues
of trapped acoustic-inertial modes for an accretion disk around a Maclaurin
spheroid of ellipticity $e = 0.834 583 178$. We are using the same
equation and boundary conditions as before, Eqs.~(\ref{prime}),~(\ref{bc}),
with the functional form of $\kappa^2(r)$ and $\Omega^2(r)$
appropriate for the chosen Maclaurin spheroid. 
The only other change is that we need to reinterpret $H$:
the condition of hydrostatic equilibrium is $c_{\rm s}=h\Omega_\perp$,
 with $h$ being the half-thickness of the disk, and $\Omega_\perp$ 
the vertical epicyclic
frequency which we absorb into an effective half-thickness
$H=h\Omega_\perp(a)/\Omega(a)$.
The results are summarized in Figs.~\ref{mac0}, \ref{mac1}, \ref{mac2},
 \ref{mac3}, \ref{mac4} for modes with $m=0,1,2,3,4$, respectively.
The frequencies are compared  in Table~(\ref{eigens})
with those obtained in the previous
sections for the black-hole disk.
For both the GR (``KL'') and the Newtonian (Maclaurin) $m=0$ 
 model the ratio of the $\mu=2$ frequency to the fundamental 
is very close to 2:1.
\bigskip

\newpage
\begin{longtable}{c|c|c|c|c}
\caption{Eigenvalues of Eqs.~(\ref{prime}),~(\ref{bc})}\\
\hline
$H/a=0.001$& $\mu=0$ & $\mu=1$ &$\mu=2$  & $\mu=3$\\
\hline
&&&&\\
 KL eq.~(\ref{kl}), $m=0$. $\tilde\sigma_\mu=$
 & 0.098829 & 0.172137 & 0.204909 & 0.226912 \\
Maclaurin,\,  $m=0$. $\tilde\sigma_\mu=$
 &  0.120977 & 0.210409 & 0.250446 &  0.277417 \\
&&&&\\
 KL eq.~(\ref{kl}), $m=1$. $\tilde\sigma_\mu=$
 & -0.880519  & -0.787151  & -0.735844 & -0.697637 \\
Maclaurin,\,  $m=1$. $\tilde\sigma_\mu=$
 & -0.576951  & -0.478477  & -0.429558 & -0.394588  \\
&&&&\\
 KL eq.~(\ref{kl}), $m=2$. $\tilde\sigma_\mu=$
 & -1.86357 & -1.753541 &  -1.688477 & -1.638747 \\
Maclaurin,\,  $m=2$. $\tilde\sigma_\mu=$
 & -1.275666  & -1.168932 & -1.112464 & -1.07094  \\
&&&&\\
 KL eq.~(\ref{kl}), $m=3$. $\tilde\sigma_\mu=$
 & -2.848839 & -2.724106 & -2.647519 & -2.588271 \\
Maclaurin,\,  $m=3$. $\tilde\sigma_\mu=$
 & -1.974981  & -1.860566  & -1.797395 & -1.750153  \\
&&&&\\
 KL eq.~(\ref{kl}), $m=4$. $\tilde\sigma_\mu=$
 & -3.83567 & -3.697564 & -3.610833 & -3.543277 \\
Maclaurin,\,  $m=4$. $\tilde\sigma_\mu=$
 & -2.674776  & -2.553138 & -2.483865 &  -2.431476 \\

\hline
\label{eigens}
\end{longtable}
This work was supported in part
by Polish NCN grant 2013/08/A/ST9/00795.

\bibliography{pkhannaRagtimeRev}

\begin{thebibliography}{26}
\expandafter\ifx\csname natexlab\endcsname\relax\def\natexlab#1{#1}\fi
\expandafter\ifx\csname url\endcsname\relax
  \def\url#1{\texttt{#1}}\fi
\expandafter\ifx\csname urlprefix\endcsname\relax\def\urlprefix{URL }\fi
\providecommand{\selectlanguage}[1]{\relax}
\providecommand{\eprint}[2][]{\url{#2}}

\bibitem[{{Abramowicz} and {Klu{\'z}niak}(2001)}]{2001A&A...374L..19A}
{Abramowicz}, M.~A. and {Klu{\'z}niak}, W. (2001), {A precise determination of
  black hole spin in GRO J1655-40}, \emph{A\&A}, \textbf{374}, pp. L19--L20.

\bibitem[{Amsterdamski et~al.(2002)Amsterdamski, Bulik, Gondek-Rosi\'nska and
  Klu\'zniak}]{amster02}
Amsterdamski, P., Bulik, T., Gondek-Rosi\'nska, D. and Klu\'zniak, W. (2002),
  {Marginally stable orbits around Maclaurin spheroids and low-mass quark
  stars}, \emph{A\&A}, \textbf{381}, p. L21.

\bibitem[{Binney and Tremaine(1987)}]{binney87}
Binney, J. and Tremaine, S. (1987), {Galactic Dynamics}, \emph{Princeton
  University Press}, p. 359.

\bibitem[{Friedman and Schutz(1978)}]{friedman78}
Friedman, J.~L. and Schutz, B.~F. (1978), Lagrangian perturbation theory of
  nonrelativistic fluids, \emph{ApJ}, \textbf{221}, p. 937.

\bibitem[{Gondek-Rosi\'nska et~al.(2014)Gondek-Rosi\'nska, Klu\'zniak,
  Stergioulas and Wi\'sniewicz}]{mateusz2014}
Gondek-Rosi\'nska, D., Klu\'zniak, W., Stergioulas, N. and Wi\'sniewicz, M.
  (2014), {Epicyclic frequencies for rotating strange quark stars: Importance
  of stellar oblateness}, \emph{Phys.~Rev.~D}, \textbf{89}, p. j4001.

\bibitem[{Kato(1989)}]{kato89}
Kato, S. (1989), Low-frequency, one-armed corrugation waves in relativistic
  accretion disks, \emph{PASJ}, \textbf{41}, p. 745.

\bibitem[{Kato and Fukue(1980)}]{fukato1980}
Kato, S. and Fukue, J. (1980), Trapped radial oscillations of gaseous disks
  around a black hole, \emph{PASJ}, \textbf{32}, p. 377.

\bibitem[{Kato et~al.(1998)Kato, Fukue and Mineshige}]{bluebook}
Kato, S., Fukue, J. and Mineshige, S. (1998), {Black-Hole Accretion Disks},
  \emph{Kyoto University Press}.

\bibitem[{{Klu{\'z}niak} et~al.(2004){Klu{\'z}niak}, {Abramowicz} and
  {Lee}}]{2004AIPC..714..379K}
{Klu{\'z}niak}, W., {Abramowicz}, M.~A. and {Lee}, W.~H. (2004),
  {High-frequency QPOs as a problem in physics: non-linear resonance}, in
  P.~{Kaaret}, F.~K. {Lamb} and J.~H. {Swank}, editors, \emph{X-ray Timing
  2003: Rossi and Beyond}, volume 714 of \emph{American Institute of Physics
  Conference Series}, pp. 379--382, \eprint{astro-ph/0402013}.

\bibitem[{Klu\'zniak et~al.(2001)Klu\'zniak, Bulik and
  Gondek-Rosi\'nska}]{kbg-r01}
Klu\'zniak, W., Bulik, T. and Gondek-Rosi\'nska, D. (2001), Quark stars in
  low-mass x-ray binaries: for and against, \emph{Proceedings of the Fourth
  INTEGRAL Workshop, Ed: B. Battrick}, \textbf{ESASP 459}, p. 301.

\bibitem[{{Klu{\'z}niak} et~al.(2005){Klu{\'z}niak}, {Lasota}, {Abramowicz} and
  {Warner}}]{2005A&A...440L..25K}
{Klu{\'z}niak}, W., {Lasota}, J.-P., {Abramowicz}, M.~A. and {Warner}, B.
  (2005), {QPOs in cataclysmic variables and in X-ray binaries}, \emph{A\&A},
  \textbf{440}, pp. L25--L28.

\bibitem[{Klu\'zniak and Lee(2002)}]{kl2002}
Klu\'zniak, W. and Lee, W.~H. (2002), The swallowing of a quark star by a black
  hole, \emph{MNRAS}, \textbf{335}, p. L29.

\bibitem[{Klu\'zniak and Rosi\'nska(2013)}]{krosinska2013}
Klu\'zniak, W. and Rosi\'nska, D. (2013), {Orbital and epicyclic frequencies of
  Maclaurin spheroids}, \emph{MNRAS}, \textbf{434}, p. 2825.

\bibitem[{Mishra and Vaidya(2014)}]{mishra2014}
Mishra, B. and Vaidya, B. (2014), {Geometrically thin accretion disk around
  Maclaurin spheroid}, \emph{A\&A}, p. submitted.

\bibitem[{Nowak and Wagoner(1991)}]{nowak1991}
Nowak, M.~A. and Wagoner, R.~V. (1991), {Diskoseismology: Probing accretion
  disks. I - Trapped adiabatic oscillations}, \emph{ApJ}, \textbf{378}, p. 656.

\bibitem[{Nowak and Wagoner(1992)}]{nowak1992}
Nowak, M.~A. and Wagoner, R.~V. (1992), {Diskoseismology: Probing accretion
  disks. II - G-modes, gravitational radiation reaction, and viscosity},
  \emph{ApJ}, \textbf{393}, p. 697.

\bibitem[{Okazaki et~al.(1987)Okazaki, Kato and Fukue}]{okazaki1987}
Okazaki, A.~T., Kato, S. and Fukue, J. (1987), Global trapped oscillations of
  relativistic accretion disks, \emph{PASJ}, \textbf{39}, p. 457.

\bibitem[{Paczy\'nski and Wiita(1980)}]{PW1980}
Paczy\'nski, B. and Wiita, P.~J. (1980), Thick accretion disks and superluminal
  luminosities, \emph{A\&A}, \textbf{88}, p.~23.

\bibitem[{Perez et~al.(1997)Perez, Silbergleit, Wagoner and Lehr}]{bob1997}
Perez, C., Silbergleit, A., Wagoner, R. and Lehr, D. (1997), {Relativistic
  Diskoseismology. I. Analytical Results for ``Gravity Modes''}, \emph{ApJ},
  \textbf{476}, p. 589.

\bibitem[{Shakura and Sunyaev(1973)}]{ss73}
Shakura, N.~I. and Sunyaev, R.~A. (1973), {Black holes in binary systems.
  Observational appearance}, \emph{A\&A}, \textbf{24}, p. 337.

\bibitem[{Silbergleit et~al.(2001)Silbergleit, Wagoner and
  Ortega-Rodr\'iguez}]{silber01}
Silbergleit, A., Wagoner, R. and Ortega-Rodr\'iguez, M. (2001), {Relativistic
  Diskoseismology. II. Analytical Results for C-modes}, \emph{ApJ},
  \textbf{548}, p. 335.

\bibitem[{{T{\"o}r{\"o}k} et~al.(2005){T{\"o}r{\"o}k}, {Abramowicz},
  {Klu{\'z}niak} and {Stuchl{\'{\i}}k}}]{2005A&A...436....1T}
{T{\"o}r{\"o}k}, G., {Abramowicz}, M.~A., {Klu{\'z}niak}, W. and
  {Stuchl{\'{\i}}k}, Z. (2005), {The orbital resonance model for twin peak kHz
  quasi periodic oscillations in microquasars}, \emph{A\&A}, \textbf{436}, pp.
  1--8.

\bibitem[{van~der Klis~M.(2000)}]{vdk00}
van~der Klis~M. (2000), {Millisecond Oscillations in X-ray Binaries},
  \emph{AnnRevA\&A,}, \textbf{38}, p. 717.

\bibitem[{Wagoner et~al.(2001)Wagoner, Silbergleit and
  Ortega-Rodr\'iguez}]{bob2001}
Wagoner, R., Silbergleit, A. and Ortega-Rodr\'iguez, M. (2001), {``Stable''
  Quasi-periodic Oscillations and Black Hole Properties from Diskoseismology},
  \emph{ApJ}, \textbf{559}, p. L25.

\bibitem[{Woudt and Warner(2002)}]{woudt02}
Woudt, P.~A. and Warner, B. (2002), {Dwarf nova oscillations and quasi-periodic
  oscillations in cataclysmic variables - I. Observations of VW Hyi},
  \emph{MNRAS}, \textbf{333}, p. 411.

\bibitem[{{Zdunik} and {Gourgoulhon}(2001)}]{zdunik01}
{Zdunik}, J.~L. and {Gourgoulhon}, E. (2001), {Small strange stars and
  marginally stable orbit in Newtonian theory}, \emph{Phys. Rev. D}.

\end{thebibliography}

\end{document}